\documentclass[aps, prb, twocolumn, showpacs, floatfix,10pt,superscriptaddress]{revtex4-2}
\usepackage{amsmath}
\usepackage{amsfonts}
\usepackage{amsbsy}
\usepackage{amssymb}
\usepackage{graphicx}
\usepackage{textcomp}
\usepackage[caption=false,singlelinecheck=false]{subfig}
\usepackage{xcolor}
\usepackage{mathrsfs}
\usepackage{mathtools}
\usepackage{bm}
\usepackage{gensymb}
\usepackage{braket,wasysym}
\usepackage[colorlinks,linkcolor=blue,citecolor=red,filecolor=magenta,urlcolor=red,breaklinks]{hyperref}
\usepackage[bottom]{footmisc}
\usepackage{verbatim}

\hypersetup{colorlinks=true, urlcolor=blue, citecolor=red, pdfborder={0 0 0}}
\usepackage{breakurl}
\usepackage{natbib}

\allowdisplaybreaks

\normalfont
\def\be{\begin{equation}}
	\def\ee{\end{equation}}
\def\bea{\begin{eqnarray}}
	\def\eea{\end{eqnarray}}

\begin{document}
\title{Length scale formation in the Landau levels of quasicrystals}

\author{Junmo Jeon}
\email{junmo1996@kaist.ac.kr}
\affiliation{Korea Advanced Institute of Science and  Technology, Daejeon 34141, South Korea}
\author{Moon Jip Park}
\email{moonjippark@ibs.re.kr}
\affiliation{Korea Advanced Institute of Science and  Technology, Daejeon 34141, South Korea}
\affiliation{Center of Theoretical Physics of Complex Systems, Institute for Basic Science (IBS) Daejeon 34126, Republic of Korea}
\author{SungBin Lee}
\email{sungbin@kaist.ac.kr}
\affiliation{Korea Advanced Institute of Science and  Technology, Daejeon 34141, South Korea}

\date{\today}
\begin{abstract}
Exotic tiling patterns of quasicrystals have motivated extensive studies of quantum phenomena such as critical states and phasons.  
Nevertheless, the systematic understanding of the Landau levels of quasicrystals in the presence of the magnetic field has not been established yet. One of the main obstacles is the complication of the quasiperiodic tilings without periodic length scales, thus it has been thought that the system cannot possess any universal features of the Landau levels.  
In this paper, contrary to these assertions, we develop a generic theory of the Landau levels for quasicrystals.
Focusing on the two dimensional quasicrystals with rotational symmetries, we highlight that quasiperiodic tilings induce anomalous Landau levels where electrons are localized near the rotational symmetry centers. 
Interestingly, the localization length of these Landau levels has a universal dependence on {$\mathbf{n}$} for quasicrystals with  $\mathbf{n}$-fold rotational symmetry. Furthermore, macroscopically degenerate zero energy Landau levels are  present due to the chiral symmetry of the rhombic tilings. In this case, each Landau level forms an independent island where electrons are trapped at given fields, but with field control, the interference between the islands gives rise to an abrupt change in the local density of states. 
Our work provide a general scheme to understand the electron localization behavior of the Landau levels in quasicrystals.
\end{abstract}
\maketitle

\textit{\textbf{Introduction---}} Quasicrystals (QC), long-range crystalline order without translational symmetry, motivates the intensive search for its unique physical phenomena. \cite{cohen2016fundamentals,kellendonk2015mathematics,baake2017aperiodic,kawazoe2003structure,suck2013quasicrystals,hauser1986magnetic,poon1992electronic,vardeny2013optics,steinhardt1987physics,janot2012quasicrystals,wang1987two,kohmoto1987electronic,divincenzo1999quasicrystals}. Of particular interest is the unique rotational symmetries that are forbidden in the conventional crystalline solid. Whereas the crystalline systems in two dimension only permit the two-,three-,four-, and six-fold rotational symmetries, the quasicrystals can have arbitrary $\mathbf{n}$-fold rotational symmetry at the cost of losing the lattice translation symmetry. Although the lack of translational symmetry invalidates the applications of the conventional Bloch theorem, alternatively, interesting wave function properties that are not observable in the conventional crystalline systems have been investigated\cite{kohmoto1987electronic,mace2017critical}. 
Along with this, unconventional transport properties and the magnetic response have been experimentally observed in these systems. In addition, exotic magnetism and superconductivity have been subsequently observed in these quasicrystalline systems\cite{goldman2014magnetism,kraus2012topological,chen2020higher,hua2020higher,varjas2019topological,zeng2020topological,huang2019comparison,barrows2019emergent}. 

The quasicrystals in the presence of the external magnetic field can exhibit highly non-trivial Landau levels (LL) since the wave functions cannot be described by the competition between the Fermi wavelength and the magnetic lengths. In this work, we study the electron localization properties of the LL in general $\mathbf{n}$-fold rotational symmetric quasicrystals with rhombic tilings. We find that the Aharonov-Bohm type destructive interference of the LL leads to the strictly localized states of the electron wave functions. This phenomenon has been similarly studied in the context of Aharonov-Bohm cage in crystalline systems\cite{vidal1998aharonov,mukherjee2018experimental}. However, for quasiperiodic case, exotic tiling patterns make a unique formation of islands for electron localization and possible interference between the islands via the control of the external magnetic field.

\begin{figure}[]
	\centering
	\includegraphics[width=0.45\textwidth]{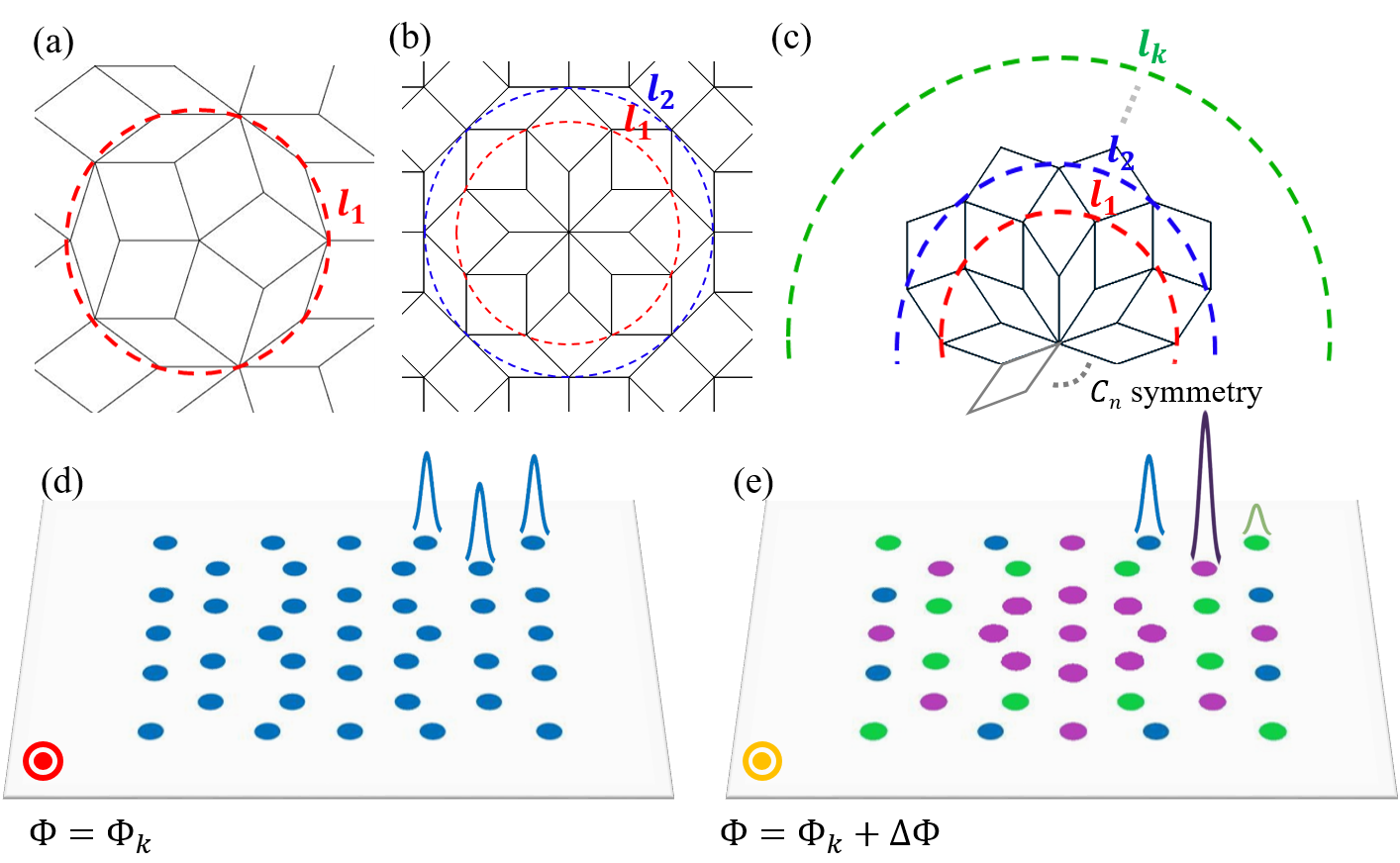}
	\label{fig1}
	
	\caption{\label{fig:scheme} (Color online) Schematic figure of the localization radius for the Landau levels emerged in (a) Penrose tiling ($n=5$), (b) Ammann-Beenker tiling ($n=8$), and (c) general $n$-fold rotational symmetric quasicrystals with rhombic tiling. The Landau levels for strict electron localization with localization radius, $l_k$, emerge at the flux, $\Phi_k$. (d) The wave functions of the localized Landau levels arise near the local rotation symmetry centers of the quasicrystals. Hence they form independent islands illustrated by the circles. (e) When the flux deviates from $\Phi_k$, the interference between localized islands enhances the localization. The different colored circles imply different amplitude of the wave functions on each island.
	}
\end{figure}
\begin{table*}[htp]
	
	\begin{center}
		\begin{tabular}{cccccccccc}
			\hline\hline
			& Penrose\cite{austin2005penrose} & Ammann-Beenker\cite{beenker1982algebraic} & Unodecagonal\cite{grunbaum1987tilings,schoen} & Heptadecagonal\cite{grunbaum1987tilings,schoen} & General  \\
			\hline
			Order of rotations ($n$)  & 5  & 8    & 11     & 17     &$ n$           \\
			Number of localization length & 1      & 2          & 4      & 7              & $\frac{n-3}{2}$ ($\frac{n}{2}-2$) \\
			\hline\hline
		\end{tabular}
	\end{center}
	\caption{Summary of the number of the possible localization lengths, $l_k$, as a function of the order of rotations, $n$. For example, in eight-fold ($n=8$) rotational symmetric Ammann-Beenker tiling, the two localization lengths, $l_1$ and $l_2$, exist. For general $n$, the number of possible localization lengths is $\frac{(n-3)}{2}$ ($\frac{n}{2}-2$) when $n$ is an odd (even) integer.}
	\label{table1}
\end{table*}

Our key results show that the LL wave functions for rotational symmetric QCs with rhombic tiling are strictly localized within certain radius, $l_k$, at the rotation centers of the QCs as shown in Fig.\ref{fig:scheme}. 
And this radius $l_k$ is only dependent on $n$ for the $n$-fold rotational symmetric QCs as,
\begin{align}
	\label{eq:mainresults}
	l_k=l\frac{\sin \pi (k+1)/n }{\sin \pi /n},
\end{align}
where $l$ is the unit length of the rhombus of the QC. $k$ is an integer which labels distinct localization lengths and magnetic fluxes. 
Depending on the order of rotations, $n$, the number of localization lengths is uniquely determined as shown in Fig.\ref{fig:scheme} (a)-(c) and Table.\ref{table1}. 
The well-known Penrose tiling and Ammann-Beenker tiling also belong to this category with $n=5$ and $n=8$ respectively as shown in Fig.\ref{fig:scheme} (a) and (b). Moreover, we show that the effective chiral symmetry of the rhombic tiling of QCs ensures the macroscopically degenerate zero-energy LLs, where strictly localized electrons form independent islands near the rotational symmetric centers of QCs. With magnetic field control, the interference between these islands occurs resulting in an abrupt change in the local density of states. (See Fig.\ref{fig:scheme} (d) and (e))

\begin{figure}[]
\centering
\includegraphics[width=0.45 \textwidth]{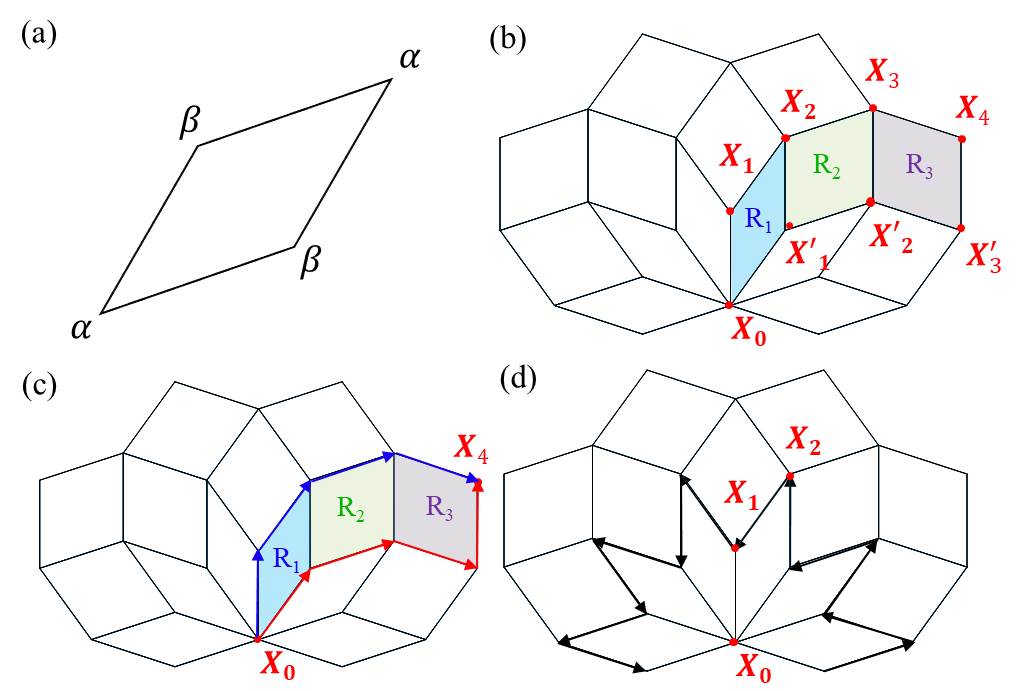}
\caption{\label{fig3} (a) Bipartition of the sites in rhombus. Each rhombus in the QCs consists of the four sites, which one can separate into two groups, $G_\alpha$, and $G_\beta$. In the tight-binding model, the sites in $G_\alpha$ can only hop to the sites in $G_\beta$ and vice versa. (b) Classification of the local site and the rhombus in generic $n$-fold rotational symmetric QC. $X_0$-class sites are defined as the local rotational symmetric centers. $X_i$-class sites are $i$-th distant sites from the rotation symmetric centers. We define $R_k$-type rhombus as $k$-th distant rhombus from the rotation symmetry center. (c) The paths connecting in between $X_0$-class site and $X_4$-class site exist  in pair (red and blue arrow lines). The combination of those two paths encircles $R_1$,$R_2$,$R_3$-type rhombus exactly once. (d) Loop consists of the class $X_1$ and $X_2$ sites (black arrow lines).
}
\end{figure}

\textit{\textbf{Generic quasicrystals---}} We start our discussion by illustrating generic $n$-fold rotational symmetric QCs with rhombic tiling. The tight-binding model is comprised of the uniform hopping terms, $\sum_{i,j}t| i \rangle\langle j| $, where $i$ and $j$ indicate the two sites connected by the edge of rhombus in the tiling. Fig. \ref{fig3} (b) shows the typical $n$-fold rotation symmetry centers of the tiling with the classifications of the local sites. Here, $X_0$-class sites are defined as the local rotation symmetry centers. $X_i$-class sites with $i=1,2, \cdots$ are classified in order of the distance from $X_0$-class sites (See Fig. \ref{fig3} (b)). We can also classify different types of the rhombus. $R_1$-type rhombus is the rhombus surrounding $X_0$-class sites and $R_k$-type rhombus with $k>1$ is defined the rhombus in order of the distance from $X_0$-class sites. Now, we consider the application of uniform magnetic field perpendicular to the two-dimensional plane by applying the Peierls substitution. {It turns out that, at certain fluxes $\Phi_k$ defined in Eq.\eqref{eq:flux}, the strictly localized LLs emerge where the wave functions perfectly vanish beyond $X_{k+1}$-class sites.}

Such strict localization of the LLs can be understood by considering the Aharonov-Bohm interference at the rotation  symmetry centers. Let's consider the wave function initially localized only at $X_0$-class sites. Then, under the time evolution, the wave function propagates along the paths that connect $X_0$-class sites and $X_k$-class sites. Their paths exist in pairs such that the combination of the two paths encircles different types of rhombuses exactly once (For exmaple, Red and blue lines in Fig. \ref{fig3} (c) encircle $R_0$,$R_1$,...,$R_{k-1}$-type rhombus). Therefore, the perfect destructive interference between the pair of the paths occurs when the flux,  $\Phi_k$, is given as,
\bea
\Phi_k&=&(2N+1)\pi\frac{1}{\sum_{i=1}^{k}\mathcal{A}_i}
\\
\nonumber &=&(2N+1)\pi\frac{\sin{\frac{\pi}{n}}}{l^2\sin{k\frac{\pi}{n}}\sin{\frac{\pi}{n}(k+1)}}.
\label{eq:flux}
\eea
Here, $\mathcal{A}_i$ is the area of the $R_{i}$-type rhombus and $k$ is a positive integer smaller than $\frac{n-2}{2}$ (See Supplementary material Sec.1 for the detailed calculations). 
For the flux $\Phi_k$, the electrons are perfectly localized in between $X_0$ class sites and $X_k$ class sites with the localization radius $l_k$ defined in Eq. \eqref{eq:mainresults}. 

One can use this analysis to find the general condition for the destructive interference for any $k$ values.  
Table \ref{table1} summarizes the number of the possible localization lengths for several examples of the well-known QCs. For larger $n$, there are larger number of localization length scales.
For example, we find that the Penrose tiling ($n=5$) only has a single localization length, but for the eight-fold ($n=8$) rotational symmetric Ammann-Beenker tiling, the two different localization radius $l_1$ and $l_2$ exist (See Fig.\ref{fig:scheme} (a) and (b)).
For Unodecagonal ($n=11$) and heptadecagonal ($n=17$) rhombic QCs, on the other hand, 11 and 17 number of distinct localization length scales exist respectively. In generic $n$-fold rotational symmetry, we prove that $n$-fold rotational symmetric QC can host $\frac{(n-3)}{2}$ ($\frac{n}{2}-2$), when $n$ is an odd (even) integer (See Supplementary Materials Sec.2 for details). According to this result, it is worthwhile to point out that, for two-dimensional crystalline systems, $n\leq6$ can only have a single $l_k$, whereas, for general QCs, there can be multiple localization length scales $l_k$.

\begin{figure}[]
	\centering
	\includegraphics[width=0.5 \textwidth]{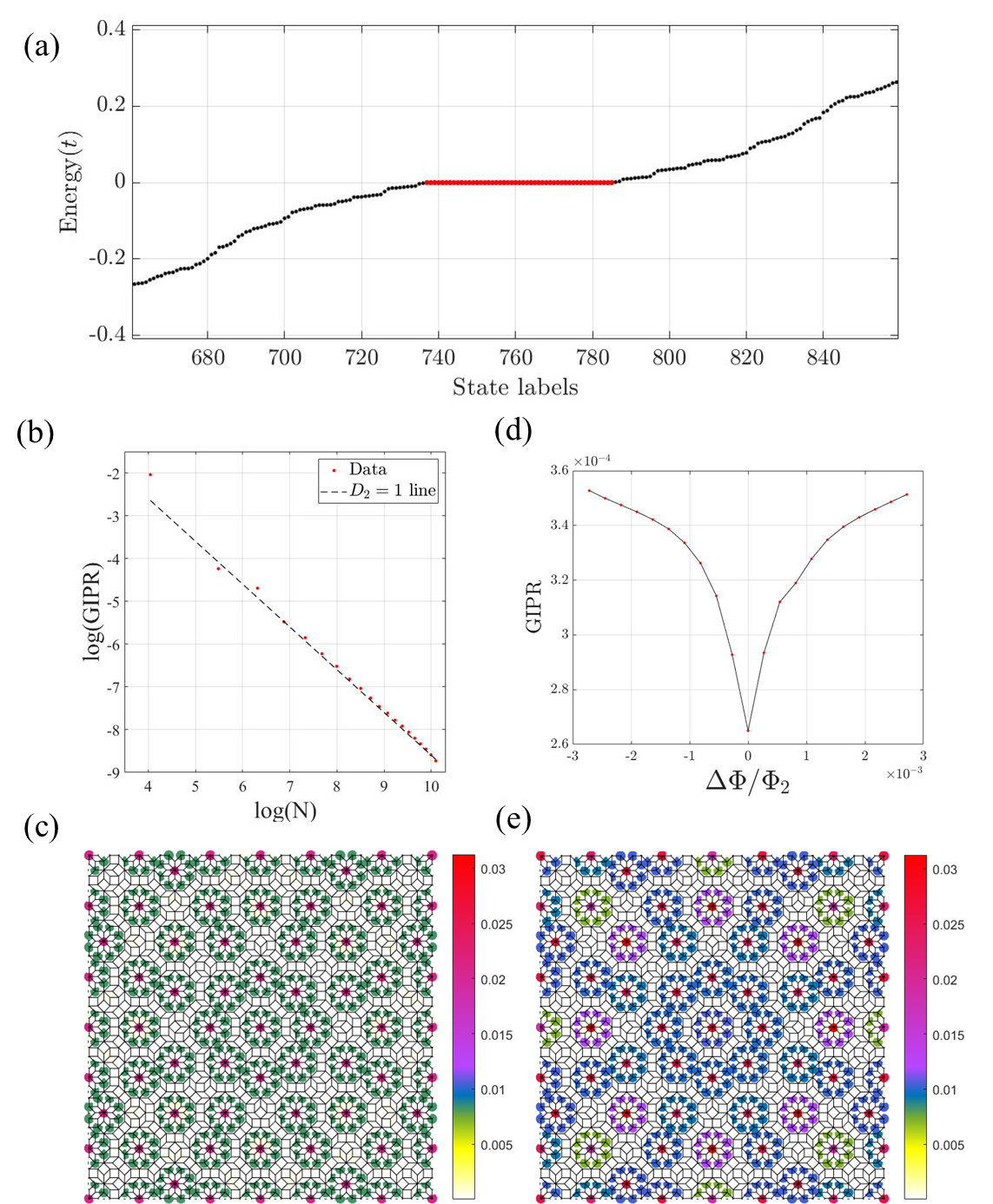}
	\caption{\label{fig4}  (a) Energy levels of the LLs at $\Phi=\Phi_2$. We find the zero energy strictly localized LLs with localization length $l_2$ (red lines).  (b) The zero-energy GIPR at $\Phi=\Phi_2$ as a function of the system size. The strictly localized LLs have the fractal dimension $D_2=1$, indicating that the wave functions are extended. For each zero energy LL, it forms an island near the rotation symmetry center and the number of such islands is macroscopically increasing with the system size. (d) GIPR as a function of the magnetic flux. 
	With magnetic flux change, the GIPR increases, indicating the enhancement of the localization. (c,e) The real-space distribution of the zero-energy LDOS at (c) $\Delta\Phi/\Phi_2=0$ and (e) $\Delta\Phi/\Phi_2=10^{-3}$. (c) The LDOS is strictly localized at $X_3$-class sites. Each rotation symmetry center has the same LDOS amplitude. (e) When the flux is varied, the strict localization fails. The interference between the islands occurs and the wave function has different LDOS amplitudes. 
	}
\end{figure}

With a given flux, $\Phi_k$, the energy levels of the strictly localized LLs show very interesting characteristics. Regardless of the specific values of $\Phi_k$, there are fixed energy levels.  In particular, the LLs at $\Phi_1$ with the localization length $l_1$ occurs at $E_1=\pm|t|\sqrt{n}$ or $E_2=0$ (See Supplementary Materials Sec.3 for the details).
For $\Phi_2$ with $l_2$, the zero energy LL with $E_2=0$ is present (See Fig. \ref{fig4} (a)).  It is important to note that the emergence of this zero-energy LL is not accidental but is originated from the chiral symmetry of the QCs which will be discussed in the next section.

To exemplify above general argument, we consider the case of the eight-fold rotational symmetric Ammann-Beenker (AB) tiling. (Without loss of generality, the same analysis can be applicable for general $n$-fold rotational symmetric QCs.) For the wave function propagation from $X_0$ class site to $X_2$ class site, there exist two paths: $X_0\to X_1\to X_2$ and $X_0\to X_1'\to X_2$. Thus, the amplitude of the wave function at $X_2$-class site is determined by the interference effect between the two paths as, $|\psi(X_2)|=t|\psi(X_0)|\left(1+\exp\left(il^2 \sin(2\pi/8)\Phi\right)\right)$, where $\Phi$ is the flux per unit area and $l$ is the length of the edge of a rhombus. The wave function at $X_2$-class sites vanishes when $\Phi=\Phi_1\equiv \frac{\left(2N+1\right)\pi}{l^2 \sin(2\pi/8)}$, where $N\in \mathbb{Z}$. As a result, the LL is strictly localized in the radius $ l_{1}={2 l \cos{\frac{\pi}{8}}}$, which is one of the simple example of Eq. \eqref{eq:mainresults}  for $n=8$. 
The wave functions of the LLs in the AB tiling can be confined within the larger distance, in between $X_0$ class sites and $X_3$-class sites. By calculating  all the possible paths that connects from $X_0$-class site to $X_3$- class sites, we find that the LLs are strictly localized within the radius $l_2=l (1+ 2\cos{ \frac{2\pi}{8}})$ at the flux, $\Phi_2 \equiv \frac{\left(2 N+1\right)\pi}{l^2 \sin(2\pi/8)}\frac{1}{(1+2\cos{\frac{2\pi}{8}})}$, where $N\in \mathbb{Z}$. 

\textit{\textbf{Zero-energy Landau levels---}} The effective chiral symmetry ensures the particle-hole symmetric energy spectrum of generic $n$-fold rotational symmetric QCs with rhombic tiling. In rhombic tiling, each rhombus consists of four sites, as shown in Fig. \ref{fig3} (a). In the tight-binding model, one can group the four sites into the two distinct groups, each site of which can only hop to the sites in the other group. We call the two groups, $G_\alpha$ and $G_\beta$ respectively, and the tight-binding model Hamiltonian can be written in the following form as,
\bea
H_{bip}=\sum_{i_\alpha \in G_\alpha , j_\beta\in G_\beta}t_{i_\alpha j_\beta}| i_\alpha \rangle\langle j_\beta|+ t^*_{i_\alpha j_\beta}| j_\beta \rangle\langle i_\alpha|,
\eea
where $t_{i_\alpha j_\beta}$ is the hopping term that connects between the $i_\alpha$-site in $G_\alpha$ and the $j_\beta$-site in $G_\beta$. In this form of the Hamiltonian, the effective chiral symmetry operator, $\Xi$, is exactly defined as, $\Xi\equiv\sum_{i_\alpha\in G_\alpha}| i_\alpha \rangle\langle i_\alpha| -\sum_{i_\beta\in G_\beta}| i_\beta \rangle\langle i_\beta|$. The Hamiltonian, $H_{bip}$, preserves the chiral symmetry by satisfying the following condition, $\{\Xi ,H\}=0$. For any eigenstate $|\psi_E \rangle$ with the energy $E$, we can define the eigenstate, $\Xi |\psi_E \rangle$, with the negative energy, $-E$, since it satisfies $H\Xi |\psi_E\rangle=-\Xi H |\psi_E\rangle=-E\Xi |\psi_E\rangle$. As a result, the energy spectrum of the LLs in the QCs with rhombic tiling is particle-hole symmetric. 

Furthermore, the chiral symmetry ensures the existence of the zero-energy  LLs at any $l_k$. Here, we only provide the sketch of the proof for {$k=2$}  and the rigorous proof for the general values of $k$ is provided in the Supplementary Material Sec.3. The Schr\"{o}dinger equation for the strictly localized LLs with the localization length $l_2$ can be written as following.

\bea
\label{linearsystem}
\begin{pmatrix}
	0 & T_{01}^\dagger &  0 \\
	T_{01}  & 0 &  T_{12}^\dagger \\
	0 & T_{12} & 0\\
\end{pmatrix}
\begin{pmatrix}
	\psi_{X_0}\\
	\psi_{X_1}\\
	\psi_{X_2}\\
\end{pmatrix}
=E
\begin{pmatrix}
	\psi_{X_0}\\
	\psi_{X_1}\\
	\psi_{X_2}\\
\end{pmatrix},
\eea
where $T_{IJ}$ represents the hopping matrix from $I$-class sites to $J$-class sites. $\psi_{X_i}$ indicates the vector of the wave functions at $X_{i}$-class sites. The chiral symmetry constrains $\psi_{X_1}=0$ but $\psi_{X_0}\neq 0$ for the zero energy wave function. Then, the linear equations in Eq. \eqref{linearsystem} host zero-energy solution if and only if the matrix $T_{12}^{\dagger}$ of the left-hand side is invertible. Using the standard Gaussian elimination method, we find that the matrix becomes non-invertible only if the following condition is satisfied,
\begin{align}
	\label{wonderful}
	1+(-1)^{(n-1)}e^{i\mathcal{A}_{\textrm{loop},2}\Phi}=0,
\end{align}
where $\mathcal{A}_{\textrm{loop},2}$ is the area of the loop that consists of $X_1$-class sites and $X_2$-class sites. The above equation is valid only if the $\Phi/\Phi_1$ is a rational number. Since the $\Phi_2/\Phi_1$ is an irrational number, the above condition is not satisfied. Therefore, the strictly localized zero-energy LLs exist. For $k\neq2$, one can also check the invertibility using the Gaussian elimination methods. 

\textit{\textbf{Macroscopic degeneracy---}}  Now let's quantify the localization behavior in generic magnetic flux. We calculate the generalized inverse participation ratio (GIPR), which is defined as, $\textrm{GIPR}(\omega)\equiv\sum_{i}\rho(i,\omega)^{2}/[\sum_{i}\rho(i,\omega)]^{2}$, where $\rho(i,\omega)$ is the local density of states at the $i$-th site with the energy $\omega$. The dependence of the system size, $N$, characterizes the multifractality of the wave functions by $\textrm{GIPR}(\omega)\propto N^{-D_2}$, with the fractal dimension $D_2$;  $D_2=0 $ corresponds to the localized states, $D_2=1$ corresponds to the extended states and $0<D_2<1$ indicates the multi-fractal critical states\cite{roche1997electronic}. Fig. \ref{fig4} (b) shows the calculated GIPR of the AB tiling at zero energy as a function of the system size. Interestingly, we find that the fractal dimension of the strictly localized LLs corresponds to $D_2=1$ i.e., the extended states. This indicates that, although the LLs are strictly localized at the rotation symmetry centers, the number of such rotation symmetry centers proportionally increases with the system size. Thus, one can imagine that the strictly localized zero energy LLs form an island near each rotation symmetry center and the macroscopic number of such islands exist as shown in Fig. \ref{fig4} (c) which is also illustrated in Fig.\ref{fig:scheme} (d).
The emergence of this macroscopic LLs is the unique feature of the rotational symmetric QCs. 

 As the magnetic flux is slightly changed from $\Phi_2$, the strict localization fails, and the wave functions of the LLs extend beyond each island. Fig. \ref{fig4} (d) shows the GIPR as a function of the magnetic field.
By slightly changing the flux  $\Phi=\Phi_2+\Delta \Phi$, the GIPR rather increases, which indicates the enhancement of the localizations, resulting in an abrupt change in local density of states (LDOS). Thus, the GIPR has local minima at $\Phi=\Phi_2$. To understand the enhancement of the localization, we compare the LDOS at $\Phi_2$ and $\Phi_2+\Delta \Phi$ (Fig. \ref{fig4} (c) and (e)). Fig. \ref{fig4} (c) shows that the identical LDOS amplitude in each rotational symmetric center. This is the consequence of strict localization of wave functions forming independent islands with $\Phi_2$. In contrast, Fig. \ref{fig4} (e) shows that each rotational center has a different LDOS amplitude. It indicates the onset of the interference between the islands that enhances the GIPR. 

\textit{\textbf{Discussions and Conclusion---}}
To summarize, we theoretically demonstrate that the anomalous strictly localized LLs on the rotational symmetric QCs with rhombic tiling. We show that non-crystallographic rotational symmetries\cite{baake2017aperiodic} generate multiple strictly localization radius, $l_k$. Furthermore, we show that the chiral symmetry of the rhombic tiling QCs gives rise to the macroscopically degenerated zero-energy LLs for any $l_k$. As a result, under the special magnetic fields, the independent islands are formed by these strictly localized electrons. By controlling the magnetic field strength, we find that the interference between the islands enhances the amount of localization in terms of increase of GIPR. We emphasize that our general works on the rhombic tiling QCs indeed illustrate the anomalous LLs that are forbidden in the conventional crystalline system as shown in Table.\ref{table1}. Furthermore, the exotic tiling pattern of QCs induces the unique formation of the islands for strictly localized LLs and their interferences as illustrated in Fig.\ref{fig:scheme}.

\subsection*{Acknowledgement}
This work is supported by National Research Foundation Grant (NRF-
2020R1F1A1073870, NRF-2020R1A4A3079707),).

\bibliography{reference}

\clearpage
\pagebreak

\renewcommand{\thesection}{\arabic{section}}
\setcounter{section}{0}
\renewcommand{\thefigure}{S\arabic{figure}}
\setcounter{figure}{0}
\renewcommand{\theequation}{S\arabic{equation}}
\setcounter{equation}{0}

\begin{widetext}
	\section*{Supplementary Material}

	\section{Derivation of $\Phi_k$ and $l_k$}
	\label{sec:1}
	In this section, we present the derivation of Eq. (1) and Eq. (2) in the main text. The destructive interference at the class $X_k$ site occurs when the flux penetrating $R_1$,$R_2$,..,$R_k$-type rhombus becomes $(2N+1)\pi$, where $N\in\mathbb{Z}$. Here, the area of the $R_j$-th rhombus is given as $\mathcal{A}_j=l^2 \sin(2j\theta_n)$ where $\theta_n=\pi/n$. The total area is given as $\sum_{j=1}^{k}l^2\sin(2j\theta_n)$. The magnetic flux across this area is $(2m_k+1)\pi$ where $m_k$ is an integer at destructive interference. Since the localization flux condition, $\Phi_k$, is defined as the flux across an unit area, we have $\Phi_k=\frac{(2m_k+1)\pi}{\sum_{j=1}^{k}\sin(2j\theta_n)}$. Using the geometric sum, we can simplify the expression of the total area of $R_1$,$R_2$,..,$R_k$-type rhombus as,
	\begin{align}
		\label{eq:phiderivation1}
		&\mathcal{A}_{\textrm{total}}=\sum_{j=1}^{k}\sin(2j\theta_n)=\mathfrak{Im}\left(e^{2i\theta_n}\frac{1-e^{2ik\theta_n}}{1-e^{2i\theta_n}} \right)=\frac{\sin{k\theta_n}\sin{(k+1)\theta_n}}{\sin{\theta_n}}.
	\end{align}
	where $\mathfrak{Im}(z)$ is the imaginary part of $z$. Hence we arrive at the result in Eq. (2) in the main text. In addition, $l_k$ is the sum of the diagonal lengths of the odd (even) numbered rhombuses. Explicitly, for integer $m$, we can write $l_k$ as,
	\begin{align}
		\label{eq:lderivation1}
		&l_{2m}=l\left(1+2\sum_{j=1}^{m}\cos{2j\theta_n}\right),
		&l_{2m+1}=2l\sum_{j=0}^{m}\cos{(2j+1)\theta_n}.
	\end{align}
	By explicitly calculating the summations in the above Eq. \eqref{eq:lderivation1}, we derive the Eq. (1) in the main text.
	\begin{align}
		\label{eq:lderivation3}
		&l_{k}=l \frac{\sin{(k+1)\theta_n}}{\sin{\theta_n}}
	\end{align}
	
	\section{Maximum number of $l_k$ for generic quasicrystals}
	\label{sec:2}
	
	In this section, we prove our Table.1 in the main text. The strict localization of the LLs fails if the tiling admits the next nearest neighbor class hoppings. For concrete argument, we illustrate a local patch of general $n$-fold rotational symmetric rhombic tiling around the $X_0$-class site, local rotational symmetric center with some examples of the rhombic tilings in Fig.\ref{fig:_rhombus}; (b) Penrose tiling (5-fold) (c) Ammann-Beenker tiling (8-fold) (d) unodecagonal tiling (11-fold) and (e) heptadecagonal tiling (17-fold) respectively. In Fig.\ref{fig:_rhombus}(a), the angles are given by $\angle X_kX_{k-1}X_k'=\frac{2k\pi}{n}$ for $n$-fold rotational symmetric rhombic tiling by using the mathematical induction. The sites in each class admit the nearest-neighbor class hoppings only when these angles are internal angles of a single rhombus. Since an internal angle of the rhombus is less than $\pi$, these angles give the desired necessary conditions to host the localization flux condition.
	\begin{figure}
		\centering
		\includegraphics[width=0.7 \textwidth]{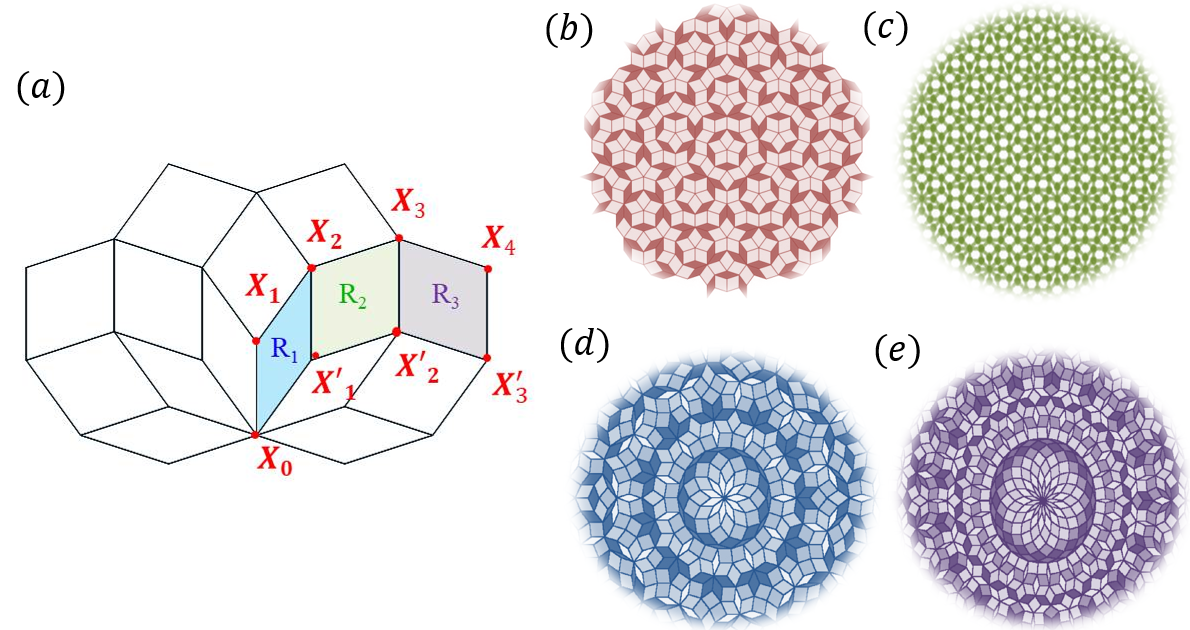}
		\caption{\label{fig:_rhombus} (a) Local patch of general $n$-fold rotational symmetric rhombic tiling around a $X_0$-class sites. We term them as $A$-class for simplicity. Similarly, we term $X_k$-class sites using $k$-th alphabet. The angles give necessary conditions of connectivities to host $\Phi_k$. The sites are classified as in our main text. Some examples of $n$-fold rotational symmetric rhombic tilings are illustrated; (b) $n=5$, (c) $n=8$, (d) $n=11$ and (e) $n=17$.}
	\end{figure}
	Thus, $\Phi_k$ arises only when $\angle X_{k+1}X_{k}X_{k+1}'<\pi$ i.e. $n>2(k+1)$. For instance, the $\Phi_1$ arises only when $n>4$. Hence there is no $\Phi_1$ in the periodic crystalline structures with $n=3$ or $4$. Note that the Penrose tiling with $n=5$ (see Fig.\ref{fig:_rhombus}(b)) satisfies this condition. Hence the Penrose tiling in Fig.\ref{fig:_rhombus}(b) hosts $\Phi_1$. Similarly, the $\Phi_2$ arises only when $n>6$. Thus, the Penrose tiling never hosts $\Phi_2$. On the other hand, the Ammann-Beenker tiling in Fig.\ref{fig:_rhombus}(c), which is $n=8$ case hosts $\Phi_2$ as we have discussed in the main text. We generalize our argument to the general types of localization flux conditions and hence get Table.1 in the main text. Particularly, for two-dimensional crystalline systems, $n\leq6$ can only have a single $l_k$, whereas, for general QCs, there can be multiple localization length scales $l_k$.

	\section{Energy levels of strictly localized Landau levels}
	\label{sec:4}
	In this section, we derive fixed energy levels, $E_1=\pm|t|\sqrt{n}$ and $E_2=0$. Especially, we prove the existence of the zero-energy strictly localized Landau level(LL)s at any $k$. First, let's derive the $E_1=\pm|t|\sqrt{n}$ for $k=1$. The Schrodinger equation of the strictily localized LLs with $l_1$ becomes
	\bea
	\label{E1}
	\begin{pmatrix}
		0 & T_{01}^\dagger \\
		T_{01}  & 0 \\
	\end{pmatrix}
	\begin{pmatrix}
		\psi_{X_0}\\
		\psi_{X_1}\\
	\end{pmatrix}
	=E
	\begin{pmatrix}
		\psi_{X_0}\\
		\psi_{X_1}\\
	\end{pmatrix}.
	\eea
	$T_{IJ}$ represents the hopping matrix from class $I$ sites to $J$ sites. $\psi_{X_i}$ indicates the vector of the wave functions at the class $X_{i}$ sites. Since $E_1\neq0$, we restrict our interest to the nonzero energy here. Thus, $\psi_{X_1}=\frac{1}{E}T_{01}\psi_{X_0}$ and hence $T_{01}^{\dagger}T_{01}\psi_{X_0}=E^2\psi_{X_0}$. Since $\psi_{X_0}\neq0$ for nontrivial solution, we have $T_{01}^{\dagger}T_{01}=E^2$. Thus, $E=E_1=\pm|t|\sqrt{n}$.
	
Now we prove the presence of the zero-energy strictly localized LLs for any $l_k$. We first prove $k=1$ case. The wave functions of the LLs vanish on $X_2$-class sites. Thus, from the chiral symmetry, we may let the zero-energy LLs wave function vanishes on $X_0$-class site. Then, the zero energy Schrodinger equation for the strictly localized wave function becomes
\bea
\label{oddk1}
T_{01}^{\dagger}\psi_{X_1}=0.
\eea
Eq.\eqref{oddk1} has an solution given by the $\psi_{X_{1j}}=t_{01_j}e^{ij\varphi}$, where $\psi_{X_{1j}}$ and $t_{01_j}$ are the $j$-th component of $\psi_{X_1}$ and $T_{01}$, respectively. Here, $e^{i\varphi}\neq1$ is a solution of the equation $x^n-1=0$. Thus, for the shifted localization flux, $\Phi=\Phi_1-\frac{\varphi}{\sin{2\theta_n}}$, the zero-energy strictly localized LLs arise within $l_1$.

Next, we consider $k=2$ case. The Schrodinger equations for the strictly localized wave functions can be generally written as,
	\bea
	\label{linearsystem2}
	\begin{pmatrix}
		0 & T_{01}^\dagger &  0 \\
		T_{01}  & 0 &  T_{12}^\dagger \\
		0 & T_{12} & 0\\
	\end{pmatrix}
	\begin{pmatrix}
		\psi_{X_0}\\
		\psi_{X_1}\\
		\psi_{X_2}\\
	\end{pmatrix}
	=E
	\begin{pmatrix}
		\psi_{X_0}\\
		\psi_{X_1}\\
		\psi_{X_2}\\
	\end{pmatrix}.
	\eea
Since for $k=2$, the wave function of the strictly localized LLs vanishes on $X_3$-class sites, the chiral symmetry allows us to set $\psi_{X_1}=0$ but $\psi_{X_0}\neq 0$ for the zero-energy case. Eqs.\eqref{linearsystem2} host (unique) nontrivial solution if the matrix $T_{12}^{\dagger}$ is invertible. Also, a square matrix is invertible if and only if its reduced row echelon form has no zero rows and hence the matrix has full rank. Thus, from the reduced row echelon form of the matrix for $T_{12}^{\dagger}$, one can show that the matrix is singular only when the following condition is satisfied,
	\begin{align}
		\label{beauty}
		1+(-1)^{(n-1)}\frac{\prod_{j=1}^{n}t_R(j)}{\prod_{j=1}^{n}t_G(j)}=0,
	\end{align}
	where $t_G$ and $t_R$ are the set of the hopping terms that are illustrated with the green and the red arrow lines respectively in Fig. \ref{figs2} .
	\begin{figure}[]
		\centering
		\includegraphics[width=0.8 \textwidth]{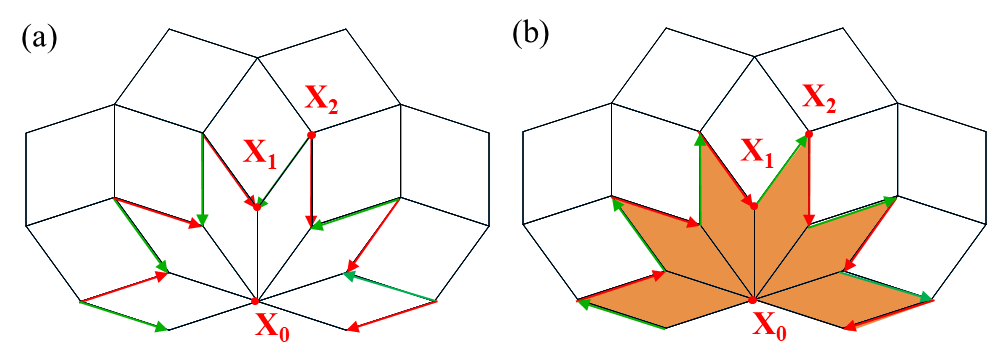}
		\caption{\label{figs2} Schematic illustration of Eq.\eqref{beauty} and Eq.\eqref{wonderful}. The arrows represent the hopping terms $t_G$ and $t_R$ as their colors green (shaded) and red, respectively. Since the hopping magnitude is uniform in the edge model, the inverse of the hopping term is equivalent to the inversed direction in the line integral of the vector potential $\bold{A}(\bold{r})$ on the phase. This fact is shown as reversed arrows in (b). The arrows enclose the orange region that gives rise to gauge invariant result. See main text for details. The directions of arrows represent that the direction of the hopping. 
		}
	\end{figure}
	By inverting the direction of the hopping terms, we can rewrite the above equation in terms of the flux enclosing a loop consists of the red and green arrow (orange area in Fig. \ref{figs2}).
	\begin{align}
		\label{wonderful}
		1+(-1)^{(n-1)}e^{in\mathcal{A}_1\Phi}=0.
	\end{align}
	Here, the $\mathcal{A}_1$ is the area of the rhombus $R_1$. The above condition is satisfied only when $n\mathcal{A}_1\Phi=(2M+n)\pi$, where $M\in\mathbb{Z}$. However, $\Phi=\Phi_2$ does not fit this condition. Therefore, the zero-energy localized Landau levels exists when $\Phi=\Phi_2$. We now generalize the proof for any even integers $k>2$. We perform the similar Gaussian elimination method. In this case, we have multiple conditions that the Hamiltonian matrix becomes singular:
	\begin{align}
		\label{evenk}
		1+(-1)^{(n-1)}e^{inA(r)\Phi}=0, \ \ 1\le r \le s
	\end{align}
	where $A(r)=\sum_{j=1}^{2r-1}\sin(2j\pi/n)=\frac{\sin{(2r-1)\theta_n}\sin{2r\theta_n}}{\sin{\theta_n}}$. Here, $\theta_n=\pi/n$. The above considtion is not satisfied when $\Phi=\Phi_{2s}$ in Eq. (2) in the main text.
	
	When $k=2s+1$ is an odd integer, we should consider the wave function which vanishes at $X_0$-class sites like the case of $k=1$. The conditions to do \textit{not} have zero-energy localized LLs are given by,
	\begin{align}
		\label{oddk}
		1+(-1)^{(n-1)}e^{inB(r)\Phi}=0, \ \ 1\le r \le s
	\end{align}
	where $B(r)=\sum_{j=1}^{2r}\sin(2j\pi/n)=\frac{\sin{2r\theta_n}\sin{(2r+1)\theta_n}}{\sin{\theta_n}}$. However, for the odd $k$ case, from the $\psi_{X_0}=0$, we have a constraint on the wave function as Eq.\eqref{oddk1}. Thus, as like the case of $k=1$, there is the \textit{shifts} of the localization flux. Again the possible shifts are given by the solutions of the equation $x^n-1=0 \ (x\neq 1)$. Hence for the phase of one of the solution of the equation $x^n-1=0 (x\neq1)$, say $\varphi$, we would have the desired zero-energy strictly localized Landau levels confined in the $l_k$ at the flux $\Phi=\Phi_k-\frac{\varphi}{S(k)}$, where $S(k)=\frac{\sin{k\theta_n}\sin{(k+1)\theta_n}}{\sin{\theta_n}}$ is the total area of $R_1,R_2,\cdots R_k$-type rhombus (See Eq.\eqref{eq:phiderivation1}). To do \textit{not} have zero-energy localized Landau levels at these shifted localiztion fluxes, $\Phi=\Phi_k-\frac{\varphi}{S(k)}$ should satisfy Eqs.\eqref{oddk}, simultaneously. However, this is not allowed because $e^{in\varphi}=1$ i.e. $nB(r)\Phi =\alpha Z\pi$ where $\alpha$ is an irrational and $Z$ is an integer. Thus, the zero-energy localized Landau levels which are confined in the $l_k$ islands arise under the shifted localized flux conditions. In conclusion, the zero-energy strictly localized LLs appear for any $l_k$.

\end{widetext}

\end{document}